\documentclass[showpacs,showkeys,twocolumn,preprintnumbers,amsmath,amssymb]{revtex4}
\newcommand{\beq}{\begin{equation}}
\newcommand{\eeq}{\end{equation}}

\newcommand{\id}{i\kern.06em\hbox{\raise.25ex\hbox{$/$}\kern-.60em$\partial$}}

\newcommand{\bs}{/\kern-.52em b}
\newcommand{\qs}{/\kern-.52em s}
\newcommand{\D}{{\cal{D}}}

\newcommand{\p}{\partial}
\newcommand{\yp}{^{\prime}}

\newcommand{\dd}
{\kern.06em\hbox{\raise.25ex\hbox{$/$}\kern-.60em$\partial$}}

\newcommand{\Tr}{\mathop{\rm Tr}\nolimits}

\newcommand{\bi}{{\bf i}}
\newcommand{\bj}{{\bf j}}

\newcommand{\bk}{{\bf k}}
\newcommand{\bp}{{\bf p}}
\newcommand{\bq}{{\bf q}}
\newcommand{\bl}{{\bf l}}

\newcommand{\JK}{J_{\rm K}}

\newcommand{\bS}{{\bf S}}

\newcommand{\bra}[1]{\langle #1|}
\newcommand{\ket}[1]{|#1\rangle}

\newcommand{\lj}{\langle}
\newcommand{\rj}{\rangle}
\DeclareMathAlphabet{\mathpzc}{OT1}{pzc}{m}{it}
\usepackage{mathrsfs}
\input amssym.def
\input amssym
\input epsf
\usepackage{graphicx}
\usepackage{dcolumn}
\usepackage{bm}
\usepackage{epsfig}

\begin{document}
\title{A low temperature derivation of spin-spin exchange in Kondo lattice model}
\author{Sze-Shiang Feng$^\dag$, Mogus Mochena}
\affiliation
 {Physics Department, Florida A \& M  University, Tallahassee, FL 32307}
\begin{abstract}
Using Hubbard-Stratonovich transformation and drone-fermion
representations for spin-$\frac{1}{2}$ and for spin-$\frac{3}{2}$,
which is presented for the first time , we  make a path-integral
formulation of the Kondo lattice model. In the case of weak
coupling and low temperature, the functional integral over
conduction fermions can be approximated to the quadratic order and
this gives the well-known RKKY interaction. In the case of strong
coupling, the same quadratic approximation leads to an effective
local spin-spin interaction linear in hopping energy $t$.
\end{abstract}
\maketitle
\section{Introduction}
In systems in which there is little direct overlapping of wave functions
of localized magnetic atoms, the spin-spin interaction mediated by itinerant electrons
is the dominant magnetic interaction, and has been an interesting topic for several decades.
In particular, the study of spin-spin interaction in diluted magnetic semiconductors(DMS) has drawn much attention
in the past several years. One of the key issues in explaining the magnetic properties of DMS is how the very
diluted holes mediate the interactions between the localized, randomly distributed $\frac{5}{2}$-spins
\cite{s1}-\cite{s5}.
When the coupling between itinerant fermions and spins is weak compared to the Fermi energy, perturbation theory applies.
Ruderman and Kittel initially considered the indirect exchange coupling of nuclear magnetic moments
by conduction electrons\cite{s6}. And later, this interaction, conventionally called RKKY exchange,
was extended by Bloembergen and Rowland\cite{s7}
, Kasuya\cite{s8} and Yosida\cite{s9}. Usually, RKKY interaction can be obtained using second order perturbation theory
\cite{s10}. But in the strong coupling regime, the conventional perturbation theory doesn't work and we need to devise
a new way to derive the effective spin-spin exchange. In section II, we present a path-integral formulation of Kondo
lattice model, in which local spins are represented by drone-fermions. Here the drone-fermion representation
of spin-$\frac{3}{2}$ is presented for the first time. At low temperatures, the spin-spin exchange is obtained.
In section III we discuss a two site system and the origin of ferromagnetic interaction linear in $t$. The last section
is a brief summary.\\
\section{The path-integral and the spin-spin exchange}
\indent  We consider Kondo lattice model with Hamiltonian
 \beq
H=\sum_{\lj\bi,\bj\rj,\sigma}t_{\bi\bj}c^\dag_{\bi\sigma}c_{\bj\sigma}+\JK\sum_\bi
\bS_\bi\cdot{\bf s}_\bi
-\mu\sum_\bi(n_{\bi\uparrow}+n_{\bi\downarrow}) \eeq where
$c^\dag_{\bi\sigma}(c_{\bi\sigma})$ denotes the
creation(annihilation) operator of the itinerant electrons (or
holes) and $n_{\bi\sigma}=c^\dag_{\bi\sigma}c_{\bi\sigma}$ is the
number operator. $\bS_\bi$ denotes the localized spins, either $S=\frac{1}{2}$
or $S=\frac{3}{2}$. ${\bf s}=\frac{1}{2}c^\dag\mbox{\boldmath$\tau$}c$ is the
electron spin where $\mbox{\boldmath$\tau$}=(\tau_1, \tau_2,
\tau_3)$ is the vector of the three usual Pauli matrices.  Hopping
energy $t_{\bi\bj}=t>0$ if $\bi,\bj$ are nearest neighbors and zero otherwise,
$\JK>0$ is the Kondo coupling and $\mu$ denotes
chemical potential. The symbol $\lj\bi,\bj\rj$ implies that the
summation in the first term is taken over nearest neighbors. The
other two summations are taken throughout the lattice.  Note that
there exists a local spin on every site and these local spins do
not interact with each other directly. To derive the mediated
interaction, we need to separate the conduction fermions and
the local spins first. One way to this is to introduce an
auxiliary field and use Hubbard-Stratonovich transformation. For
this,  one needs to express the coupling $\bS\cdot{\bf s}$ in terms
of squares of Hermitian, bounded operators. There are different
ways to do this. For example, one can write \beq \bS\cdot{\bf
s}=\frac{1}{2}[(\bS+{\bf s})^2-\bS^2-{\bf
s}^2]=-\frac{1}{2}[(\bS-{\bf s})^2-\bS^2-{\bf s}^2] \eeq Using
$\bS^2=S(S+1)$ and ${\bf
s}^2=-\frac{3}{4}(n_\uparrow+n_\downarrow)^2+\frac{3}{2}(n_\uparrow
+n_\downarrow)$ \beq \bS\cdot{\bf s}=\frac{1}{2}[(\bS+{\bf
s})^2-S(S+1)+\frac{3}{4}(n_\uparrow+n_\downarrow)^2-\frac{3}{2}(n_\uparrow
+n_\downarrow)] \eeq or \beq \bS\cdot{\bf
s}=-\frac{1}{2}[(\bS-{\bf
s})^2-S(S+1)+\frac{3}{4}(n_\uparrow+n_\downarrow)^2-\frac{3}{2}(n_\uparrow
+n_\downarrow)] \eeq To formulate the path-integral, we also need
 a representation for the local spins.
 For spin$\frac{1}{2}$,we can use the drone-fermi representation\cite{s11}:
$
S^z=f^\dag f-\frac{1}{2},\,\,\,\,\,\,\,\,\,
S^+=f^\dag(d+d^\dag),\,\,\,\,\,\,\,\,\, S^-=(d+d^\dag)f
$
where $f, d$ are fermionic operators. One of the most important conveniences of this
representation is that unlike the Schwinger fermion representation, this representation does not require constraints.
For spin-$\frac{3}{2}$, we have a generalization (to our best knowledge, this generalization is new, ref.\cite{s12}
only represents the spin operators in terms of fermi operators and the representation is obviously not
a drone-fermion representation since one term is bosonic and the other is fermionic) :
$
S^+=\sqrt{3}f^\dag_1(d+d^\dag)+2f^\dag_2 f_1,
S^-=\sqrt{3}(d+d^\dag)f_1+2f^\dag_1f_2, S^z=f^\dag_1f_1+2f^\dag_2f_2-\frac{3}{2}
$
. Note that we use 3 fermi fields for spin-$\frac{3}{2}$. In general, this type of fermi representation is viable for
$S-$spin when $2^{n-1}<2S+1\leq 2^n$, as counted in\cite{s12}. Thus it is seen that for both spin-$\frac{1}{2}$ and
spin$-\frac{3}{2}$, we can write the partition function as purely fermionic path-integral. Using expression (3) and the relation
(the so-called Hubbard-Stratonovich transformation) $e^{-A^2}=\frac{1}{\sqrt{\pi}}
\int ^\infty_{-\infty}dx e^{-(x^2+i2Ax)}$, which holds
for Hermitian and bounded operator $A$ , we can write the partition function as
\begin{widetext}
\begin{eqnarray}
\mathcal{Z}&=&\int\D\mbox{\boldmath$A$}_\bi(\tau) \D\varphi_\bi(\tau)
e^{-\int^\beta_0 d\tau \sum_\bi(\mbox{\boldmath$A$}_\bi^2+\varphi^2_\bi)}
\int\D c_{\bi\sigma}(\tau)
\D c^*_{\bi\sigma}(\tau)\int\D d_{\bi\sigma}(\tau)
\D d^*_{\bi\sigma}(\tau)\D f_{\bi\sigma}(\tau)\D f^*_{\bi\sigma}(\tau)\nonumber\\
&&
\times e^{-\int ^\beta_0 d\tau [\sum_{\bi\sigma}(c^*_{\bi\sigma}\p_\tau c_{\bi\sigma}
+f^*_{\bi\sigma}\p_\tau f_{\bi\sigma}+d^*_{\bi\sigma}\p_\tau d_{\bi\sigma})+\mathcal{H}]}
\end{eqnarray}
where $\D$ denotes functional measure and $\mbox{\boldmath$A$}_\bi(\tau), \varphi_\bi(\tau)$ are
 auxiliary random fields and the extended Hamiltonian
\begin{eqnarray}
\mathcal{H}=t\sum_{\lj \bi,\bj\rj,\sigma}c^\dag_{\bi\sigma}c_{\bj\sigma}-
(\mu+\frac{3}{4}\JK)\sum_\bi(n_{\bi\uparrow}+n_{\bi\downarrow})+\sum_\bi
[i\sqrt{2\JK}(\bS_\bi+{\bf s}_\bi)\cdot\mbox{\boldmath$A$}_\bi+i
\sqrt{\frac{3}{2}\JK}\varphi_\bi(n_{\bi\uparrow}+n_{\bi\downarrow})]
\end{eqnarray}
\end{widetext}
Using the Fourier transformation  $
\mbox{\boldmath$A$}_{m\bk}=N^{-1}\sum_\bi \beta^{-1/2}\int^\beta_0 d\tau
e^{i\omega^B_m\tau}e^{i\bk\cdot\bi}\mbox{\boldmath$A$}_\bi(\tau)$ where $
\omega^B_m=\frac{2m}{\beta}\pi, (m=0,1,..)$ is the Matsubara frequency for bosonic fields, the electron part of the functional-integral is written as
\begin{widetext}
\begin{eqnarray}
\mathcal{Z}_c[\mbox{\boldmath$A$},\varphi]&=&
\int\prod_{n\sigma} dc_{n\bk\sigma}d c^*_{n\bk\sigma} e^{-\sum_{mn,\bk,\bp}
c^*_{m\bk}[(i\omega_n+t_\bk-\mu-\frac{3}{4}\JK)\delta_{\bp,\bk}\delta_{mn}
+i\sqrt{\frac{3}{2N\beta}\JK}\varphi_{m-n,\bk-\bp}
+i\sqrt{\frac{1}{2N\beta}\JK}\mbox{\boldmath$\sigma$}\cdot\mbox{\boldmath$A$}_{m-n,\bk-\bp}]c_{n\bp}}
\end{eqnarray}
where $\omega_n=\frac{2n+1}{\beta}\pi, (n=0,1,..)$ is the Matsubara frequency for fermionic fields. Using the path-integral technique for Grassmann variable, we have
\begin{eqnarray}
\ln \mathcal{Z}_c[\mbox{\boldmath$A$},\varphi]&=&
\ln \frac{\text{Det}[(i\omega_n+t_\bk-\mu-\frac{3}{4}\JK)\delta_{\bp,\bk}\delta_{mn}
+i\sqrt{\frac{3}{2N\beta}\JK}\varphi_{m-n,\bk-\bp}
+i\sqrt{\frac{1}{2N\beta}\JK}\mbox{\boldmath$\sigma$}\cdot\mbox{\boldmath$A$}_{m-n,\bk-\bp}}
{\text{Det}(i\omega_n\delta_{mn})}\nonumber\\
&=&\ln\text{Det}[(\p_\tau+t_\bk-\mu-\frac{3}{4}\JK)\delta_{\bp,\bk}]-\ln\text{Det}(\p_\tau)\nonumber\\&&
+\ln \text{Det}[\delta_{mn}\delta_{\bk,\bp}+(i\omega_m+t_\bk-\mu-\frac{3}{4}\JK)^{-1}
(i\sqrt{\frac{3}{2N\beta}\JK}\varphi_{m-n,\bk-\bp}
+i\sqrt{\frac{\JK}{2N\beta}}
\mbox{\boldmath$\sigma$}\cdot\mbox{\boldmath$A$}_{m-n,\bk-\bp})]
\end{eqnarray}
At low temperatures such that
$\sqrt{\JK/\beta}\ll $ the largest of $t, \JK$ and $\mu$, the third term can be calculated perturbatively. Using
$\ln\text{Det}A=\Tr\ln A$, we have
\begin{eqnarray}
&&\ln \text{Det}[\delta_{mn}\delta_{\bk,\bp}+(i\omega_m+t_\bk-\mu-\frac{3}{4}\JK)^{-1}
(i\sqrt{\frac{3}{2N\beta^2}\JK}\varphi_{m-n,\bk-\bp}
+i\sqrt{\frac{\JK}{2N\beta^2}}
\mbox{\boldmath$\sigma$}\cdot\mbox{\boldmath$A$}_{m-n,\bk-\bp})]\nonumber\\
&=&\Tr[(i\omega_m+t_\bk-\mu-\frac{3}{4}\JK)^{-1}
(i\sqrt{\frac{3}{2N\beta}\JK}\varphi_{m-n,\bk-\bp}
+i\sqrt{\frac{\JK}{2N\beta}}
\mbox{\boldmath$\sigma$}\cdot\mbox{\boldmath$A$}_{m-n,\bk-\bp})]\nonumber\\&&
-\frac{1}{2}\Tr[(i\omega_m+t_\bk-\mu-\frac{3}{4}\JK)^{-1}
(i\sqrt{\frac{3}{2N\beta}\JK}\varphi_{m-n,\bk-\bp}
+i\sqrt{\frac{\JK}{2N\beta}}
\mbox{\boldmath$\sigma$}\cdot\mbox{\boldmath$A$}_{m-n,\bk-\bp})]^2+\cdots
\end{eqnarray}
To the quadratic order, i.e., neglecting the part $\cdots$, we have
\begin{eqnarray}
&&\ln \text{Det}[\delta_{mn}\delta_{\bk,\bp}+(i\omega_m+t_\bk-\mu-\frac{3}{4}\JK)^{-1}
(i\sqrt{\frac{3}{2N\beta^2}\JK}\varphi_{m-n,\bk-\bp}
+i\sqrt{\frac{\JK}{2N\beta^2}}
\mbox{\boldmath$\sigma$}\cdot\mbox{\boldmath$A$}_{m-n,\bk-\bp})]\nonumber\\&\simeq&2\sum_n\frac{i\sqrt{\frac{3}{2N\beta}\JK}\varphi_{0,0}}{i\omega_n+t_\bk-\mu-\frac{3}{4}\JK}\nonumber\\&&
-\frac{1}{2}\sum_{\bp,\bk}\sum_{m,n}\Tr\frac{i\sqrt{\frac{3}{2N\beta}\JK}\varphi_{m-n,\bk-\bp}
+i\sqrt{\frac{\JK}{2N\beta}}
\mbox{\boldmath$\sigma$}\cdot\mbox{\boldmath$A$}_{m-n,\bk-\bp}}{i\omega_m+t_\bk-\mu-\frac{3}{4}\JK}
\frac{i\sqrt{\frac{3}{2N\beta}\JK}\varphi_{n-m,\bp-\bk}
+i\sqrt{\frac{\JK}{2N\beta}}
\mbox{\boldmath$\sigma$}\cdot\mbox{\boldmath$A$}_{n-m,\bp-\bk}}{i\omega_n+t_\bp-\mu-\frac{3}{4}\JK}
\end{eqnarray}
The first term vanishes in the thermodynamic limit $N\rightarrow\infty$. Therefore the leading term is
the second term
\begin{eqnarray}
&&\sum_{\bp,\bk}\sum_{m,n}\frac{\frac{3}{2N\beta}\JK\varphi_{m-n,\bk-\bp}\varphi_{n-m,\bp-\bk}
+\frac{\JK}{2N\beta}
\cdot\mbox{\boldmath$A$}_{m-n,\bk-\bp}\cdot\mbox{\boldmath$A$}_{n-m,\bp-\bk}
}{(i\omega_m+t_\bk-\mu-\frac{3}{4}\JK)(i\omega_n+t_\bp-\mu-\frac{3}{4}\JK)}\nonumber\\
&=&-\frac{\JK}{2t}\sum_{m,\bk}(3\varphi_{m,\bk}\varphi_{-m,-\bk}+
\mbox{\boldmath$A$}_{m,\bk}\cdot\mbox{\boldmath$A$}_{-m,-\bk})Q_{m\bk}
\end{eqnarray}
where
\beq
Q_{m\bk}=-\frac{1}{\beta N}\sum_{n\bp}
\frac{t}{(i\omega_{m+n}+t_{\bk+\bp}-\mu-\frac{3}{4}\JK)(i\omega_n+t_\bp-\mu-\frac{3}{4}\JK)}
\eeq
so the partition function
\begin{eqnarray}
\mathcal{Z}&=&
\int\D f_{\bi\sigma}(\tau)\D f^*_{\bi\sigma}(\tau)\D d_{\bi\sigma}(\tau)\D d^*_{\bi\sigma}(\tau)
e^{-\int ^\beta_0 d\tau \sum_{\bi\sigma}(f^*_{\bi\sigma}\p_\tau f_{\bi\sigma}
+d^*_{\bi\sigma}\p_\tau d_{\bi\sigma}
)}
\int d\mbox{\boldmath$A$}_{m\bk} d\varphi_{n\bk}
e^{-\sum_{m\bk}(\mbox{\boldmath$A$}_{m\bk}\cdot\mbox{\boldmath$A$}^*_{m,\bk}+\varphi_{m\bk}\varphi^*_{m,\bk})}\nonumber\\&&
\times \exp\{-
\frac{\JK}{2t}\sum_{m,\bk}(3\varphi_{m,\bk}\varphi^*_{m,\bk}+
\mbox{\boldmath$A$}_{m,\bk}\cdot\mbox{\boldmath$A$}^*_{m,\bk})Q_{m\bk}
\}e^{-\sqrt{2\JK}\sum_{m\bk} \bS_{m\bk}\cdot \mbox{\boldmath$A$}^*_{m,\bk}}
\end{eqnarray}
where $d\mbox{\boldmath$A$}_{m\bk} d\varphi_{n\bk}$ are just ordinary integration measures as
$\mbox{\boldmath$A$}_{m\bk},\varphi_{n\bk}$ are now Fourier components. We have used replacement
$A_{m\bk}\rightarrow iA_{m\bk},A^*_{m\bk}\rightarrow -iA_{m\bk}$, i.e., redefine
the Fourier component such that
$$
\mbox{\boldmath$A$}_{m\bk}=\frac{1}{N}\sum_\bi \frac{1}{\sqrt{\beta}}\int^\beta_0 d\tau
e^{i\omega^B_m\tau}e^{i\bk\cdot\bi}i\mbox{\boldmath$A$}_\bi(\tau)
$$
$$
\mbox{\boldmath$A$}_{-m,-\bk}=\frac{1}{N}\sum_\bi \frac{1}{\sqrt{\beta}}\int^\beta_0 d\tau
e^{-i\omega^B_m\tau}e^{-i\bk\cdot\bi}(-i)\mbox{\boldmath$A$}_\bi(\tau)
$$
Hence the induced spin-spin interaction comes from the integral over $\mbox{\boldmath$A$}$,
which gives
\beq
\int\D\mbox{\boldmath$A$}_{m\bk}
e^{-\sum_{m\bk}\mbox{\boldmath$A$}_{m\bk}\cdot\mbox{\boldmath$A$}^*_{m,\bk}
-\frac{\JK}{2t}\sum_{m,\bk}\mbox{\boldmath$A$}_{m,\bk}\cdot\mbox{\boldmath$A$}^*_{m,\bk}Q_{m\bk}
- \sqrt{\frac{\JK}{2}}\sum_{m\bk} (\bS_{m\bk}\cdot \mbox{\boldmath$A$}^*_{m,\bk}
+\mbox{\boldmath$A$}_{m,\bk}\cdot\bS_{m\bk}^*)}
=e^{\frac{\JK}{2}\sum_{m,\bk}\mbox{\boldmath$S$}^*_{m\bk}\frac{1}{1+\frac{\JK}{2t}Q_{m\bk}}\cdot\mbox{\boldmath$S$}_{m\bk}}
\eeq
Now since
\begin{eqnarray}
Q_{m\bk}
&=&-\frac{1}{\beta}\frac{1}{N}\sum_\bp\frac{t}{t_\bp-t_{\bk+\bp}-i\omega^B_m}\sum_n
[\frac{1}{i\omega_{m+n}+t_{\bk+\bp}-\mu-\frac{3}{4}\JK}-\frac{1}{i\omega_n+t_\bp-\mu-\frac{3}{4}\JK}]
\end{eqnarray}
\end{widetext}
, using the Matsubara frequency sum $\sum_{n=-\infty}^{\infty}\frac{1}{i\omega_n+x}=\frac{\beta}{e^{-\beta x}+1}$,
we have
\beq
Q_{m\bk}=\frac{t}{N}\sum_\bp\frac{f(t_{\bk+\bp}-\mu-\frac{3}{4}\JK)-f(t_{\bp}-\mu-\frac{3}{4}\JK)}{t_\bp-t_{\bk+\bp}-i\omega^B_m}
\eeq
where $f(x)=\frac{1}{e^{\beta x}+1}$ is the Fermi function. In the low temperature limit, we can take $\omega^B_m\simeq0$ so $Q_{m\bk}\simeq Q_\bk$, where
\beq
Q_{\bk}\simeq\frac{t}{N}\sum_\bp\frac{f(t_{\bk+\bp}-\mu-\frac{3}{4}\JK)-f(t_{\bp}-\mu-
\frac{3}{4}\JK)}{t_\bp-t_{\bk+\bp}}
\eeq
(In the case $t_\bk=\bk^2/(2m)$, this is nothing but the function $F_3(\bq)$\cite{s13} and we reproduce the RKKY
interaction).
If $t\gg\JK$, the induced interaction is
\beq
-\int^\beta_0d\tau \sum_{\bi,\bj}F(\bi-\bj)\bS_\bi\cdot\bS_\bj\simeq-\frac{\JK^2}{4t}\sum_{m,\bk}
\mbox{\boldmath$S$}^*_{m\bk}Q_{\bk}\cdot\mbox{\boldmath$S$}_{m\bk}
\eeq
which is RKKY-like since it is proportional to $\JK^2$ and inversely to $t$ which corresponds to Fermi energy.
 \\
 \indent In the strong coupling case as in diluted magnetic semiconductors,
  $t\ll \JK$. Keeping only the
  quadratic term of the functional determinant
 in (10),  the resulting spin-spin effective action at low temperatures is of the form
\beq
t\sum_{m,\bk}\mbox{\boldmath$S$}^*_{m\bk}(\frac{t}{2\JK}+Q_{\bk})^{-1}
\mbox{\boldmath$S$}_{m\bk}=-\int^\beta_0 d\tau (-t) \sum_{\bi,\bj}J_{\bi,\bj}\bS_\bi\cdot\bS_\bj
\eeq
if $Q_\bk\gg \frac{2t}{\JK}$ for all $\bk$ ,
where $J_{\bi,\bj}$ is the Fourier transform of the $(\frac{2t}{\JK}+Q_\bk)^{-1}$. This interaction is linear
in $t$. The linearity in $t$ of the induced spin-spin interaction is intuitive since when
$\JK$ is very large, then $\JK$ is irrelevant and the only energy scale left is $t$. A more formal discussion
of induced spin-spin interactions for a two site sysytem is given in the next section.
\section{Perturbative derivation of moment-moment exchange}
(1){\it Exchange between classical moments at half-filling}\\\\
\indent We now consider a half-filled lattice with a classical moment $S{\bf n}$ on each site,
where ${\bf n}$ is a unit vector.  The relation of this system
to the Kondo lattice model described by Hamiltonian (1) is as follows.
 The quantum-mechanical states of the
local spin $\bS$ can be expressed in terms of the eigenstate $\ket{{\bf n}}$ of $\bS, \bS\ket{{\bf n}}
=S{\bf n}\ket{{\bf n}}$.
We can then obtain the energy levels by considering different orientations of ${\bf n}$.
In the strong-coupling limit, the hopping term is taken as a perturbation while the on-site
interaction is the unperturbed Hamiltonian.
Using the canonical transformation
\begin{eqnarray}
c_{\bi\uparrow}&=&\frac{1}{\sqrt{2-2n_{\bi z}}}
(n_{\bi -}\alpha_\bi+(n_{\bi z}-1)\beta_\bi)\\
c_{\bi\downarrow}&=&\frac{1}{\sqrt{2-2n_{\bi z}}}
((1-n_{\bi z})\alpha_\bi+n_{\bi +}\beta_\bi)
\end{eqnarray}
where $n_{\bi\pm}=n_{\bi x}\pm in_{\bi y}$, and $\alpha_\bi, \beta_\bi$ are new fermi  fields, the on-site term becomes
 \beq
  \JK S\sum_{\bi} {\bf
s}_{\bi}(\tau)\cdot{\bf n}_{\bi}(\tau)=\frac{1}{2}\JK S\sum_{\bi}
[\alpha^\dag_{\bi}(\tau)\alpha_{\bi}(\tau)-\beta^\dag_{\bi}(\tau)\beta_{\bi}(\tau)]
\eeq
Hence at half-filling (the number of particles equals the number of lattice sites)
, the unperturbed ground-state is (for $\JK>0$)
 \beq \ket{\rm Gnd}=\prod_\bi \beta^\dag_\bi\ket{0}
 \eeq with energy $ E_{\rm Gnd}=-\frac{1}{2} \JK S N$. $N$ is the number of lattice sites.
Since the hopping part reads $H\yp=\sum_{\lj\bi,\bj\rj}H_{\bi\bj}$ where
\begin{widetext}
\begin{eqnarray}
H_{\bi\bj}=t_{\bi\bj}\frac{1}{2}\frac{1}{\sqrt{(1-n_{\bi
z})(1-n_{\bj z})}}&&
\{[n_{\bi-}n_{\bj+}+(1-n_{\bi z})(1-n_{\bj z})]\alpha^\dag_\bi\alpha_\bj
+[n_{\bi-}(n_{\bj z}-1)+(1-n_{\bi z})n_{\bj-}]\alpha^\dag_\bi\beta_\bj\nonumber\\&&
+[(n_{\bi z}-1)n_{\bj+}+n_{\bi+}(1-n_{\bj z})]\beta^\dag_\bi\alpha_\bj
+[(n_{\bi z}-1)(n_{\bj z}-1)+n_{\bi+}n_{\bj-}]\beta^\dag_\bi\beta_\bj\}
\end{eqnarray}
 the only non-vanishing term in  $H\yp\ket{\rm Gnd}$ is,
\beq
\sum_{\bi\bj}t_{\bi\bj}\frac{1}{2}\frac{1}{\sqrt{(1-n_{\bi z})(1-n_{\bj z})}}
[n_{\bi-}(n_{\bj z}-1)+(1-n_{\bi z})n_{\bj-}]\alpha^\dag_\bi\beta_\bj\ket{\rm Gnd}
\eeq
Defining states
\beq
\ket{\bi\bj}=\ket{\beta^\dag_1,...\beta^\dag_\bi\alpha^\dag_\bi,...,(_\bj),...,\beta^\dag_N}
\eeq
in which site $\bj$ is empty, site $\bi$ is doubly occupied and all other sites are singly occupied by
$\beta$ quasi-particles. Therefore the second-order perturbation is
\begin{eqnarray}
\sum_{\bk\bl}\frac{\bra{{\rm Gnd}}\sum_{\lj\bi\yp\bj\yp\rj}H_{\bi\yp\bj\yp}\ket{\bk\bl}
\bra{\bk\bl}\sum_{\lj\bi\bj\rj}H_{\bi\bj}\ket{{\rm Gnd}}}{E_{\rm Gnd}-E_m}
\end{eqnarray}
Note that
\beq
\bra{\bi\bj}H\yp\ket{{\rm Gnd}}=(-1)^{\bi+\bj}t_{\bi\bj}\frac{1}{2}\frac{1}{\sqrt{(1-n_{\bi z})(1-n_{\bj z})}}
[n_{\bi-}(n_{\bj z}-1)+(1-n_{\bi z})n_{\bj-}]
\eeq
\beq
\bra{{\rm Gnd}}H\yp\ket{m_{\bi\bj}}=(-1)^{\bi+\bj}t_{\bi\bj}\frac{1}{2}\frac{1}{\sqrt{(1-n_{\bi z})(1-n_{\bj z})}}
[n_{\bi+}(n_{\bj z}-1)+(1-n_{\bi z})n_{\bj+}]
\eeq
we get the second-order correction to the ground-state energy
\beq
\Delta E^{(2)}_{\rm Gnd}=-\sum_{\bi\bj}\frac{t^2_{\bi\bj}}{4\JK S}(2-2n_{\bi z}n_{\bj z}-n_{\bj-}n_{\bi+}-n_{\bi-}n_{\bj+})
=\sum_{\bi\bj}\frac{t^2_{\bi\bj}}{2\JK S}(\frac{{\bf S}_\bi\cdot{\bf S}_\bj}{S^2}-1)
\eeq
We see that this is an antiferromagnetic interaction,
which agrees with references \cite{s14} and \cite{s15}. For {\it less than half-filling}, the ground-state
is highly degenerate and we must use degenerate perturbation theory, which is in fact not practical since
the degeneracy is often too high to be handled. Eq.(3) indicates that the effective spin-spin exchange which is
linear in $t$ obtained in section (II) applies only to systems away from half-filling.\\\\
(2){\it Two sites energy-levels}\\\\
In the case of two sites, the Hamiltonian (1) is
\beq
H=\left(\begin{matrix}c^\dag_{1\uparrow}& c^\dag_{1\downarrow}& c^\dag_{2\uparrow}& c^\dag_{2\downarrow}\end{matrix}\right)
\left(\begin{matrix}\frac{\JK}{2}S^z_1 & \frac{\JK}{2}S^-_{1}&t&0\cr
\frac{\JK}{2}S^+_1 & -\frac{\JK}{2}S^z_1&0&t\cr
t&0&\frac{\JK}{2}S^z_{2}& \frac{\JK}{2}S^-_2\cr
0&t&\frac{\JK}{2}S^+_2& -\frac{\JK}{2}S^z_2
\end{matrix}\right)
\left(\begin{matrix}c_{1\uparrow}\cr c_{1\downarrow}\cr c_{2\uparrow}\cr c_{2\downarrow}\end{matrix}\right)
\eeq
with eigenvalues
\beq
\pm\frac{1}{2}\sqrt{4t^2+\JK^2S^2\pm2\JK St\sqrt{2+2{\bf n}_1\cdot{\bf n}_2}}
\eeq
For $\JK S\gg 2t$, they are
\begin{eqnarray}
E_1&=&-\frac{\JK S}{2}-\frac{|t|}{2}\sqrt{2+2{\bf n}_1\cdot{\bf n}_2}-\frac{t^2}{\JK S}[1-\frac{1}{4}(2+2{\bf n}_1\cdot{\bf n}_2)]\\
E_2&=&-\frac{\JK S}{2}+\frac{|t|}{2}\sqrt{2+2{\bf n}_1\cdot{\bf n}_2}-\frac{t^2}{\JK S}[1-\frac{1}{4}(2+2{\bf n}_1\cdot{\bf n}_2)]\\
E_3&=&\frac{\JK S}{2}-\frac{|t|}{2}\sqrt{2+2{\bf n}_1\cdot{\bf n}_2}+\frac{t^2}{\JK S}[1-\frac{1}{4}(2+2{\bf n}_1\cdot{\bf n}_2)]\\
E_4&=&\frac{\JK S}{2}+\frac{|t|}{2}\sqrt{2+2{\bf n}_1\cdot{\bf n}_2}+\frac{t^2}{\JK S}[1-\frac{1}{4}(2+2{\bf n}_1\cdot{\bf n}_2)]
\end{eqnarray}
\end{widetext}
If there is only one electron, only $E_1$ is filled in the ground-state with energy
\beq
E_0=-\frac{\JK S}{2}-\frac{|t|}{2}\sqrt{2+2{\bf n}_1\cdot{\bf n}_2}-\frac{t^2}{\JK S}
[1-\frac{1}{4}(2+2{\bf n}_1\cdot{\bf n}_2)]
\eeq
and the correction due to hopping is
\beq
\Delta E_0\simeq-\frac{|t|}{2}\sqrt{2+2{\bf n}_1\cdot{\bf n}_2}
\eeq
which is always ferromagnetic. When there are two electrons, i.e., the system is half-filled, $E_1, E_2$ are
filled in the ground-state and the energy is
\beq
E=E_1+E_2=-\JK S-\frac{t^2}{\JK S}(1-{\bf n}_1\cdot{\bf n}_2)
\eeq
So the correction due to hopping is
\beq
\Delta E_0=-\frac{t^2}{\JK S}(1-{\bf n}_1\cdot{\bf n}_2)
=\sum_{\bi,\bj}\frac{t^2_{\bi\bj}}{2\JK S}({\bf n}_\bi\cdot{\bf n}_\bj-1)
\eeq
This agrees with the conclusion (30) for a general half-filled lattice. It is seen from (38) and (40) that whether
the induced moment-moment interaction is linear or quadratic in $t$ depends on the number of particles.
This justifies our previous argument.\\
\section{Conclusions}
 In this paper, we present a path-integral formulation for Kondo lattice model with localized spins ($S=\frac{1}{2}$ or
 $S=\frac{3}{2}$). In both cases, the system can be described in terms of fermions conveniently since spin operators
 are difficult to deal with in either path-integral or Feynman diagram calculations.
 In the low-temperature limit, the conduction fermions can be integrated perturbatively.
and the resulting effective spin-spin exchange is RKKY-like interaction in the weak-coupling regime.
In the strong-coupling regime, the effective spin-spin exchange is linear in $t$ if $Q_\bk\gg \frac{2t}{\JK}$ for all
$\bk$ where $Q_\bk$ is the function $F_3(\bk)$\cite{s13} in the lattice case. Since $Q(\bk)$ depends on $\mu$ which is
the chemical potential, we see qualitatively that whether or not the leading order of the induced spin-spin interaction is
linear in $t$ depends on the particle density in the system. At half-filling, the leading term of interaction is
antiferromagnetic and quadratic in $t$ according to quantum-mechanical perturbation theory. It should be noted that though
 the path-integral formalism agrees with quantum-mechanical perturbative analysis qualitatively, the former does not
 show that half-filling is of criticality. Therefore further study on the path-integral
 formalism is still called for.\\
\indent This work is supported in part by the Army High Performance Computing Research
Center(AHPCRC) under the auspices of the Department of the Army, Army Research Laboratory (ARL) under
Cooperative Agreement number DAAD19-01-2-0014.\\
$^\dag$Corresponding author, e-mail: shixiang.feng@famu.edu

\end{document}